\documentclass[%
 reprint,
superscriptaddress,
 amsmath,amssymb,
 aps,
showkeys
]{revtex4-2}

\usepackage{graphicx}% Include figure files
\usepackage{dcolumn}% Align table columns on decimal point
\usepackage{bm}% bold math
\usepackage{float}
\usepackage{amsmath}  
\usepackage{siunitx}
\usepackage[hidelinks]{hyperref}
  {
      
  }
\begin{document}

\preprint{APS/123-QED}

\title{An Algorithm for Subtraction of Doublet Emission Lines in \\ Angle-Resolved Photoemission Spectroscopy}% Force line breaks with \\
%\thanks{??}%

\author{Yaoju Tarn}
 \email{yt493@cornell.edu}%Lines break automatically or can be forced with \\
    \affiliation{Department of Physics, Laboratory of Atomic and Solid State Physics, Cornell University, Ithaca, NY, 14853, USA}
 \author{Mekhola Sinha}
    \affiliation{Department of Chemistry, Johns Hopkins University, Baltimore, MD, 21218, USA}
    %\affiliation{Platform for the Accelerated Realization, Analysis, and Discovery of Interface Materials (PARADIM), Cornell University, Ithaca, NY, 14853}
 \author{Christopher Pasco}
    \affiliation{Department of Chemistry, Johns Hopkins University, Baltimore, MD, 21218, USA}
    %\affiliation{Platform for the Accelerated Realization, Analysis, and Discovery of Interface Materials (PARADIM), Cornell University, Ithaca, NY, 14853}
 \author{Darrell G. Schlom}
    \affiliation{Department of Materials Science and Engineering, Cornell University, Ithaca, NY, 14853, USA}
    %\affiliation{Kavli Institute at Cornell for Nanoscale Science, Cornell University, Ithaca, NY, 14853, USA}
    \affiliation{Platform for the Accelerated Realization, Analysis, and Discovery of Interface Materials (PARADIM), Cornell University, Ithaca, NY, 14853}
 \author{Tyrel M. McQueen}
    \affiliation{Department of Chemistry, Johns Hopkins University, Baltimore, MD, 21218, USA}
    \affiliation{Platform for the Accelerated Realization, Analysis, and Discovery of Interface Materials (PARADIM), Cornell University, Ithaca, NY, 14853}
\author{Kyle M. Shen}%
    \affiliation{Department of Physics, Laboratory of Atomic and Solid State Physics, Cornell University, Ithaca, NY, 14853, USA}
    %\affiliation{Kavli Institute at Cornell for Nanoscale Science, Cornell University, Ithaca, NY, 14853, USA}
    %\affiliation{Platform for the Accelerated Realization, Analysis, and Discovery of Interface Materials (PARADIM), Cornell University, Ithaca, NY, 14853}
\author{Brendan D. Faeth}
 \email{bdf53@cornell.edu}
    \affiliation{Department of Physics, Laboratory of Atomic and Solid State Physics, Cornell University, Ithaca, NY, 14853, USA}
    \affiliation{Platform for the Accelerated Realization, Analysis, and Discovery of Interface Materials (PARADIM), Cornell University, Ithaca, NY, 14853}


\begin{abstract}
Plasma discharge lamps are widely utilized in the practice of angle-resolved photoemission spectroscopy (ARPES) experiments as narrow-linewidth ultraviolet photon sources. However, many emission lines such as Ar-I, Ne-I, and Ne-II have closely spaced doublet emission lines, which result in superimposed replica on the measured ARPES spectra. Here, we present a simple method for subtracting the contribution of these doublet emission lines from photoemission spectra. Benchmarking against ARPES spectra of well-characterized 2D materials, we demonstrate that this algorithm manages to subtract the doublet signal and reproduce the key features of the monochromated He-I$\alpha$ spectra in a physically sound manner that reliably reproduces quantifiable dispersion relations and quasiparticle lifetimes.
\end{abstract}

%Use showkeys class option if keyword
\maketitle

%\tableofcontents

\section{\label{sec:level1}Introduction}

Angle-resolved photoemission spectroscopy (ARPES) is a powerful tool in modern experimental condensed matter physics for characterizing the electronic structure of quantum materials \cite{RMP2021_ARPES_QuantumMaterials,RMP2003_ARPES_Cuprate,RMP_metal_insulator}. A key component of any ARPES system is the light source, the characteristics of which determine much of the system's experimental capabilities. The beam profile, photon polarization, and incident flux of the photon source determines the effective spatial resolution, orbital sensitivity, and photoemission signal intensity, respectively.  Meanwhile, the photon energy $h\nu$ serves as a critical parameter in determining the probed out-of-plane momentum $k_{\perp}$.   It is for this reason that high-intensity, continuously tunable synchrotron radiation sources are strongly preferred for band mapping of 3D materials. In contrast, lab-based ARPES systems frequently rely on noble gas discharge lamps which produce a series of intense narrow emission lines at discrete photon energies depending on the gas plasma used. The most common emission lines are He-I$\alpha$ at 21.2 eV and He-II$\alpha$ at 40.8 eV.

Other gases can also be used to generate useful discharge lines at 8.4 eV (Xe I)~\cite{doi:10.1063/1.4766962,doi:10.1063/1.2818806}, 10 eV (Kr-I)~\cite{doi:10.1063/1.3488367}, 11.8 eV (Ar-I)~\cite{doi:10.1063/1.2945641}, 16.8 eV (Ne-I), and 26.9 eV (Ne-II).  
Unfortunately, many of these discharge spectra include secondary ``doublet" lines of comparable intensity and with narrow energy spacings \cite{DoubletLines} less than the ~1.5 eV that a typical lab-based vacuum ultraviolet (VUV) monochromator is able to effectively filter, resulting in a shifted, overlapping signal in the measured photoemission spectra.  

%The discharge spectra are generated by photoemission measurements of the Fermi edge of polycrystalline Au [cite supplemental].

\begin{figure*}
\includegraphics[width=\textwidth]{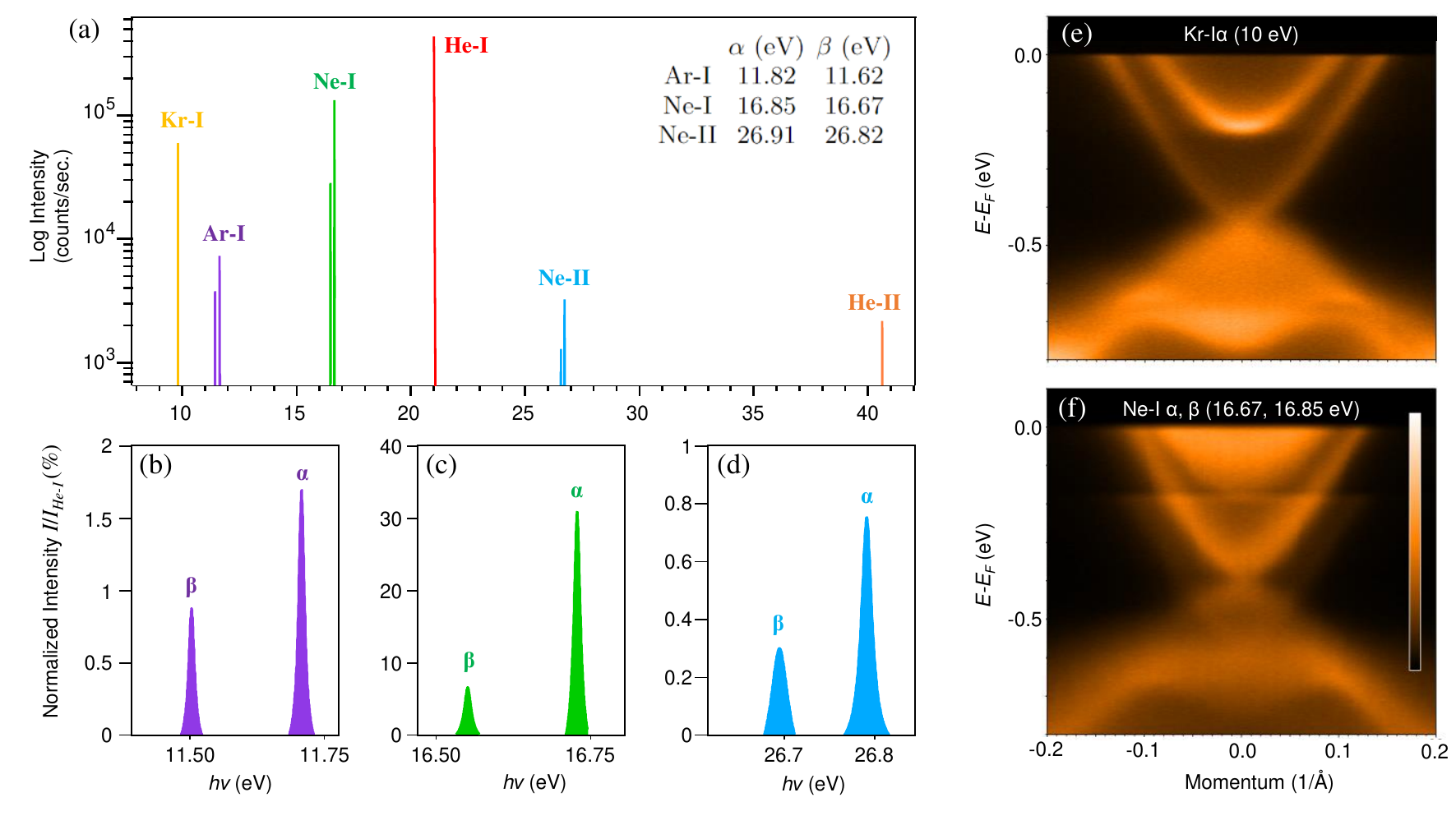}
\caption{\label{fig:problem}(a) Discharge spectra of Kr, He, Ne, and Ar gases after filtering through a Scienta Omicron VUV 5040 toroidal monochromator. Spectra are generated by taking the derivative of the angle-integrated Fermi Edge of a poly-crystalline Au sample.  See Appendix A for additional details. (b)-(d) Magnified view of the Ar-I, Ne-I, and Ne-II emission lines, normalized to the intensity of He-I. (e)-(f) ARPES spectra of Bi$_2$Se$_3$ obtained using Kr and Ne-I gases, respectively. The Kr discharge is passed through a CaF filter to produce a discrete 10 eV source, whereas the Ne-I light is imperfectly selected with the monochromator and retains signal from both Ne-I$\alpha$ and Ne-I$\beta$.
}
\end{figure*}

The challenge of using such doublet emission lines for ARPES experiments is illustrated in Fig.~\ref{fig:problem}. Our experimental system consists of a high resolution hemispherical electron analyzer (Scienta Omicron DA30-L), a toroidal monochromator (Scienta Omicron VUV 5040), and a  multi-gas RF discharge lamp (Fermion Instruments BL1200s). In Fig.~\ref{fig:problem}(a), we show VUV discharge spectra measured with this system from a polycrystalline gold reference sample using Kr, Ar, He, and Ne plasma.  The peak intensity profiles shown are generated by taking the derivative of an angle-integrated Fermi edge collected at 6 K.  Even with the monochromator, substantial secondary lines at lower $h\nu$ remain for Ar-I [Fig.~\ref{fig:problem}(b)], Ne-I [Fig.~\ref{fig:problem}(c)], and Ne-II [Fig.~\ref{fig:problem}(d)].  The effect of these secondary lines on measured ARPES spectra is clearly demonstrated in Fig.~\ref{fig:problem}(e) and Fig.~\ref{fig:problem}(f), which show unprocessed ARPES spectra of a cleaved Bi$_2$Se$_3$ single crystal at $\Gamma$ \cite{Nature2009_Bi2Se3} using both monochromated Kr-I$\alpha$ 10eV photons [Fig.~\ref{fig:problem}(e)] and Ne-I doublet emission [Fig.~\ref{fig:problem}(f)].  In the latter case, the measured spectra exhibits a pair of overlapping, duplicate copies, with the weaker Ne-I$\beta$ signal appearing at lower kinetic energy.  The presence of this overlapping doublet signal is highly problematic, particularly when attempting to extract quantitative parameters such as dispersion relations and quasiparticle lifetimes from ARPES spectra.

Consequently, there is a need for a method to eliminate the $\beta$ line contribution from ARPES spectra collected using doublet emission sources. In this work, we present a simple method for performing such a subtraction based on a piece-wise reconstruction algorithm.   %We demonstrate that []

\section{\label{sec:level1}Subtracting the Doublet Photoemission Signal}

In a conventional photoemission experiment using a monochromatic photon source with energy $h\nu$, the relationship between the measured kinetic energy $\omega$ of the outgoing electrons and the original binding energy $E_B$ is

\small
\begin{equation}
    \omega = h\nu - \phi_A - E_{B} ,
\end{equation}
\normalsize

where $\phi_A$ is the known work function of the electron analyzer.  When a doublet photon source is used, the measured photoemission signal $I_{tot}$ becomes a superposition of the emission from both lines, such that

\small
\begin{equation}
    I_{tot}(\mathbf{k},\omega) = \alpha(\mathbf{k},\omega) + \beta(\mathbf{k},\omega) ,
\end{equation}
\normalsize

where $\alpha(\mathbf{k},\omega)$ and $\beta(\mathbf{k},\omega)$ are the $\alpha$ and $\beta$ signal's contribution to the total intensity at the measured kinetic energy $\omega$ and momentum $\mathbf{k}$.  

%For a pair of doublet lines with an energy separation $\Delta E$ = $h \nu_\alpha$ - $h \nu_\beta$, it can be seen from eqs. 1 and 2 that the resultant photoemission signal at a given $\omega$ is a combination of , with the latter shifted to higher binding energy

%  We can impose an assumption with which to simplify the calculation, namely, that the doublet is almost an exact duplicate, just shifted by some amount $\Delta E$ and scaled by some factor $R$ respectively.

%By taking the energy relative to the Fermi level of the first spectrum, then assuming the same matrix elements and $k$, the entire band shifts as an approximate replica. This shift by $\Delta E$ is exemplified by the second Fermi edge visible in Fig.~\ref{fig:problem}(f). The higher energy doublet is the $\alpha$ signal while the lower energy doublet is the $\beta$ signal.

We now consider a simplified scenario, in which the measured spectral intensity scales directly with the photon flux for a given discharge line, but that the binding-energy dependent lineshapes are identical.  In this situation, the overall signal $I_{tot}(\omega)$ at a fixed $\mathbf{k}$ can be written as  

\small
\begin{equation}
    I_{tot}(\omega) = \alpha(\omega) + \beta(\omega) = \alpha(\omega) + R \alpha(\omega+\Delta E) ,
\end{equation}
\normalsize

where $R$ is simply the fractional intensity ratio of the $\beta$ photon flux to that of the more intense $\alpha$ line, and $\Delta E$ = $h \nu_\alpha$ - $h \nu_\beta$ is the doublet energy separation.  

%[Although this simplified treatment ignores the potential complicating effects of , we find that it is sufficient as a practical describing  ] 

\begin{figure*}
\includegraphics[width=\textwidth]{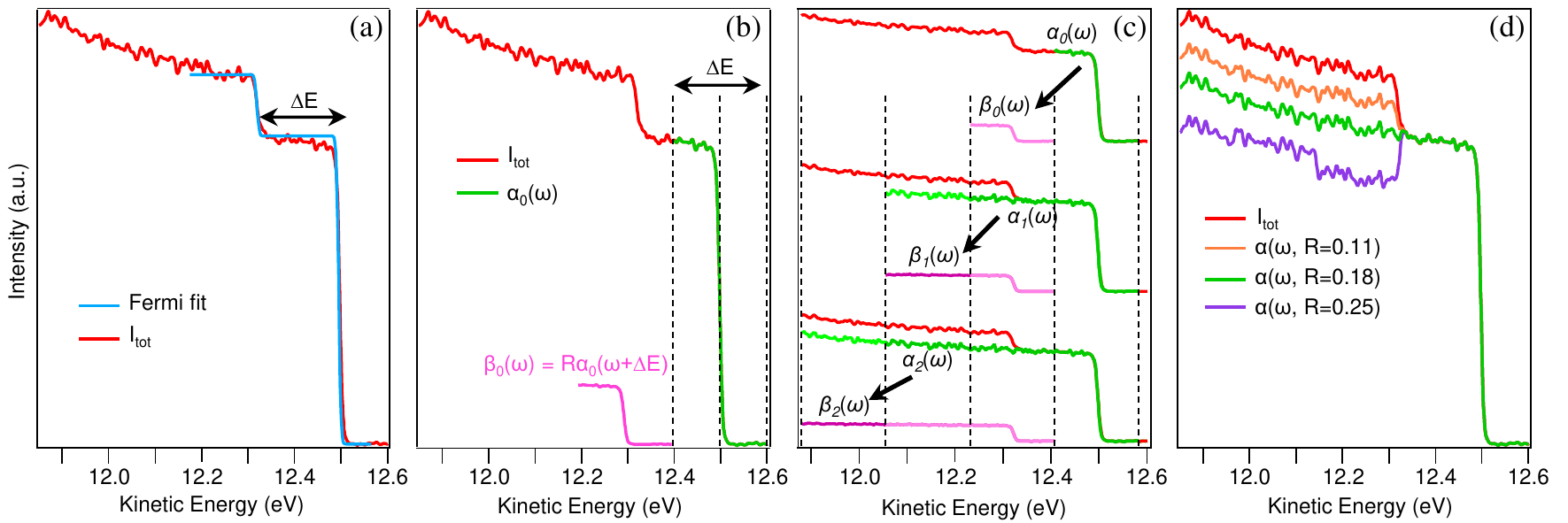}
\caption{\label{fig:flowchart}Flowchart depicting how the full doublet subtraction method works, applied on polycrystalline copper measured with a Ne-I doublet as an example. (a) $\Delta E$ is extracted by a double Fermi edge fit on the integrated EDC. (b) $\alpha_0(\omega)$ (green curve) is initialized as the segment of the raw data (red curve) spanning a window of width $\Delta E$, and $\beta_0(\omega)$ (pink curve) is initialized as a shifted and scaled copy of $\alpha_0(\omega)$. (c) $\beta_0(\omega)$ is subtracted from $\alpha_0(\omega)$ to generate $\alpha_1(\omega)$. The subtraction is repeated until it has covered the full range of the EDC. (d) The value of $R$ is optimized by minimizing the discontinuity which appears at the boundary between the first and second segments, as detailed further in Appendix B. The optimal solution (green) has this discontinuity eliminated, while a non-optimal solutions (purple and orange) continues to exhibit a discontinuity at $\Delta E$ below $E_F$.}
\end{figure*}

%FOLLOWING PARAGRAPH MAYBE NEEDS SOME CLEANUP.  
A solution for the isolated $\alpha(\omega)$ signal can be determined algorithmically from raw ARPES spectra using the form of Eq. 3 so long as $\Delta E$ and $R$ are known accurately.  We show this process diagrammatically in Fig.~\ref{fig:flowchart}. We start by accounting for background above the Fermi level as well as detector non-linearity \cite{det_nonlin} in the ARPES spectra and extract their energy distribution curves (EDCs). $\Delta E$ is determined using a Fermi edge fit [Fig.~\ref{fig:flowchart}(a)]. $\Delta E$ is a property of the chosen noble gas discharge line and does not change between experiments. In the case of Fig.~\ref{fig:flowchart},  Ne-I has a $\Delta E$ of 176 meV.

%Next, we assume that the $\beta$ signal contributes nothing above the Fermi edge of the $\alpha$ signal, i.e. the region above $E_F$ is a pure state of the $\alpha$ signal. 

Next, we define a sub-range of the original EDC in a small window $[E_F - \frac{\Delta E}{2}, E_F + \frac{\Delta E}{2}]$ straddling the Fermi level of the $\alpha$ signal [Fig.~\ref{fig:flowchart}(b)].  Within this energy range the contribution of the $\beta$ signal is negligible.  This EDC segment, denoted as $\alpha_{0}(\omega)$, is then translated by the energy separation $\Delta E$ and scaled by the intensity ratio $R$ to produce $\beta_{0}(\omega)=R\alpha_{0}(\omega+\Delta E)$, representing the binding energy equivalent segment of the doublet $\beta$ signal. The contribution of this doublet that extends into the domain of $\alpha_{0}(\omega)$ is then subtracted [Fig.~\ref{fig:flowchart}(c)], leaving $\alpha_{1}(\omega) = \alpha_{0}(\omega)-\beta_{0}(\omega)$ as a pure state of the $\alpha$ signal. This process of subtraction is subsequently repeated over the index $n$, stitching window-by-window to the domain, until it has run through the full EDC. Mathematically, the flow goes as such, with $E_F$ set to 0 for simplicity: 

\small
\begin{equation}
    \alpha_{0}(\omega) = I_{tot}(\omega), \: \omega \in [-\frac{\Delta E}{2},\frac{\Delta E}{2}]
\end{equation}

\begin{equation}
    \beta_{n}(\omega) = R \alpha_{n}(\omega+\Delta E), \: \omega \in [-(n+\frac{1}{2})\Delta E, \frac{\Delta E}{2}]
\end{equation}

\begin{equation}
    \alpha_{n\ge 1}(\omega) = I_{tot}(\omega)-\beta_{n-1}(\omega), \: \omega \in [-(n-\frac{1}{2})\Delta E, \frac{\Delta E}{2}]
\end{equation}
\normalsize

It is important to note that $R$ can vary between experiments and is affected by numerous settings such as the monochromator setting and lamp pressure. For example, with Ne-I, we find that this intensity ratio $R$ can vary over a range of $\pm$ 0.04. To account for this, we have an iterative step that varies $R$ [Fig.~\ref{fig:flowchart}(d)] to minimize the discontinuity. If $R$ is determined incorrectly, a discontinuity remains, as seen in the purple curve [Fig.~\ref{fig:flowchart}(d)]. This discontinuity is eliminated in the green curve of Fig.~\ref{fig:flowchart}(d), where the output EDCs are fully decoupled from the $\beta$ signal at the optimal $R$ determined by this minimization. The process of removing this discontinuity is discussed in detail in Appendix B.

%One challenge is that there is no residual to minimize when varying $R$. Fortunately, even a slightly incorrect $R$ of $\pm 10\%$ of the optimal $R$ value yields a visible discontinuity in the integrated EDC at $\Delta E$ below the Fermi level. This allows for the fitting of a smooth polynomial in a sub-window straddling the discontinuity, which gives a continuous, kink-less residual against which the integrated EDC of the subtraction algorithm's output can be minimized. We found empirically that choosing the width of this straddling sub-window centered around $\Delta E$ to be $\frac{\Delta E}{2}$ works well, and should be generally applicable for avoiding nearby bands. An alternative method can be implemented utilizing smoothing followed by 2nd derivative analysis in this sub-window. We have chosen a polynomial fit to avoid compounding noise when taking derivatives. We also note that the algorithm works even in the presence of noisy data. Between $10\%$ and $100\%$ ARPES scan statistics, the optimal $R$ value determined by this method only exhibits a $\sim3\%$ fractional difference.

\section{Benchmarking Performance}

In order to characterize the performance of our algorithm, we performed ARPES measurements on the misfit layered compound (LaSe)\textsubscript{1.14}(NbSe\textsubscript{2})\textsubscript{2}, which has strong two-dimensional character and is therefore ideal for comparing measured spectra at different photon energies \cite{PRB2021_MisfitQuasi2D, https://doi.org/10.1002/adfm.202007706}. Single crystal samples were cleaved under good UHV conditions ($<$ 6 x 10\textsuperscript{-11} Torr) and measured at a temperature of 6 K with a nominal instrumental energy resolution of 6 meV.  

\begin{figure*}
\includegraphics[width=\textwidth]{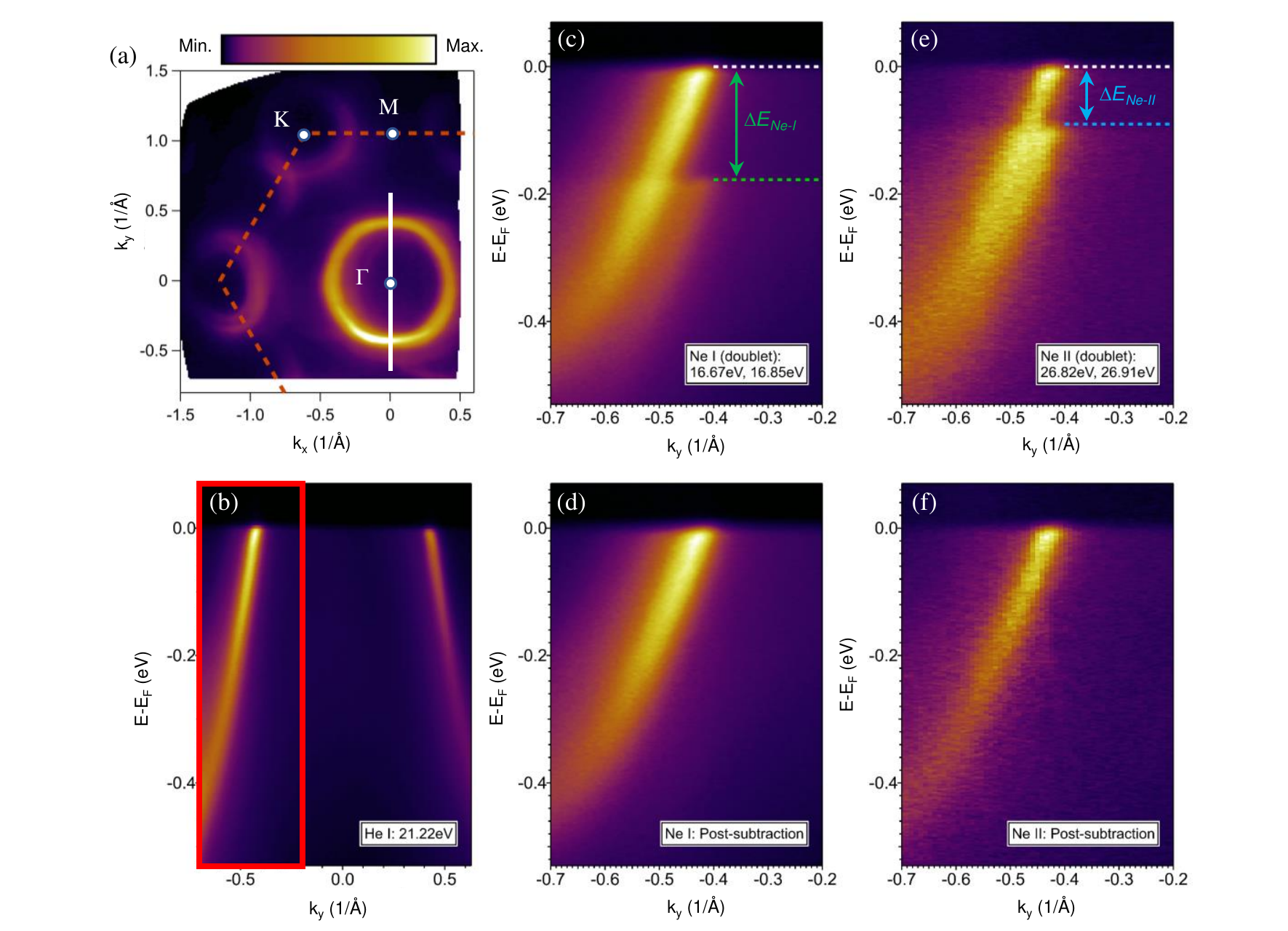}
\caption{\label{fig:misfit}(a) He-I$\alpha$ Fermi surface intensity map of the cleaved surface of (LaSe)\textsubscript{1.14}(NbSe\textsubscript{2})\textsubscript{2} single crystal, integrated over $\pm$ 5 meV of $E_F$. (b) Photoemission intensity at $\Gamma$ point [white solid line in (a)] as measured with He-I$\alpha$ radiation.  Two branches of the hole pocket are visible, with the left band at negative $k_y$ being more intense at this geometry. (c)-(d) Ne-I doublet spectra before and after subtraction, with green dotted line representing the second Fermi edge occurring at doublet separation energy of $\Delta E$ = 176 meV (e)-(f) Ne-II doublet spectra before and after subtraction, with blue dotted line representing the second Fermi edge occurring at doublet separation energy of $\Delta E$ = 90 meV.}
\end{figure*}

In Fig.~\ref{fig:misfit}, we compare ARPES measurements of (LaSe)\textsubscript{1.14}(NbSe\textsubscript{2})\textsubscript{2} using both monochromated and doublet emission sources to benchmark the efficacy of our subtraction algorithm. Fig.~\ref{fig:misfit}(a) shows the Fermi surface as measured with He-I, which consists of hole-like pockets at K and at $\Gamma$ consistent with expectation for a heavily electron-doped NbSe\textsubscript{2} surface layer as observed in previous studies~\cite{https://doi.org/10.1002/adfm.202007706,NADER1998147}. Fig.~\ref{fig:misfit}(b) shows a high-statistics energy-momentum image along the $\Gamma$-$M$ high symmetry direction of the NbSe$_2$ surface taken also with He-I$\alpha$ radiation, showing a pair of well-isolated dispersive bands crossing the Fermi level, arising from Nb 4d states of the hole-like pocket centered on $\Gamma$. 

%Figs.~\ref{fig:misfit}(c,e) show the corresponding equivalent energy-momentum images for data taken instead with Ne-I and Ne-II

%The band structure measurements obtained from the He-I spectra were used as a control against which the post-subtracted spectra were compared. (Fig.~\ref{fig:misfit}d). The isolated bright bands serve as a good platform to characterize the performance of this doublet subtraction method. 

\begin{figure*}
\includegraphics[width=\textwidth]{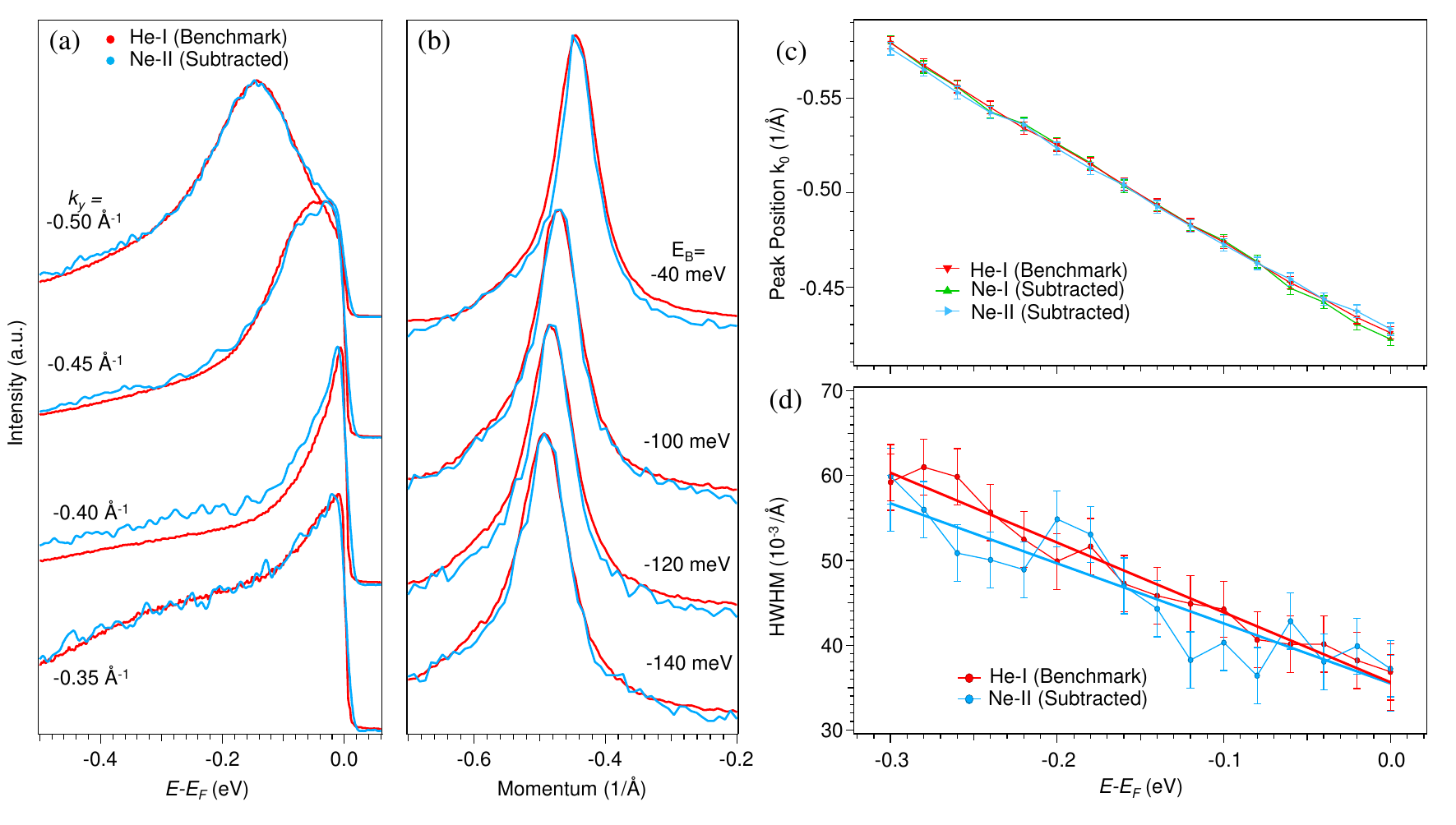}
\caption{\label{fig:quant}(a) EDCs extracted from the He-I (red lines) and post-subtracted Ne-II (blue lines) spectra of Fig.~\ref{fig:misfit}(b and f) sampled at regular momenta from -0.5 to -0.35 $\si{\angstrom}^{-1}$.  Each EDC is averaged over a momentum range of $0.04~\si{\angstrom}^{-1}$. (b) MDCs extracted from He-I and post-subtracted Ne-II spectra at several representative binding energies.  Each MDC is averaged over an energy range of 20 meV. (c) Dispersion relation (extracted from the data in Fig. 3(b,d,f). (d) MDC width analysis on Fig. 3(b) and 3(f).}
\end{figure*}

The performance of our algorithm can be benchmarked directly by comparing processed ARPES images taken using doublet emission lines against this baseline monochromated He-I spectra.  In Figures~\ref{fig:misfit}(c,e) we show raw ARPES spectra along the same momentum cut and identical measurement geometry as the He-I spectra of Fig.~\ref{fig:misfit}(b), but taken with Ne-I and Ne-II emission, respectively. For clarity, we focus here on a single band crossing the Fermi level, as indicated by the narrower momentum window shown by the red box in Fig.~\ref{fig:misfit}(b).  A second Fermi edge is visible for both raw spectra at an energy separation equal to $\Delta E$, below which the duplicate superimposed features of the beta signal are clearly evident.  In contrast, the algorithmically-subtracted doublet spectra shown in Figures~\ref{fig:misfit}(d,f) show only a single sharp band without any obvious second Fermi edge or shadow band.

%Momentum distribution curves (MDCs) were extracted at several binding energies averaged over an energy range of 200 meV and fit with Lorentzian peaks with a linear background. We used the He-I data as a benchmark [Fig.~\ref{fig:misfit}(b)] and examined the more intense band enclosed in the red box. We extracted dispersion relations and half-width half-maximums (HWHM) from the He-I, Ne-I, and Ne-II spectra after subtraction. Error bars were estimated by taking the root-mean-square of the residual in the linear portions of the dispersion relations.

%No aberrations were visible in the post-subtracted spectra, implying discontinuities in the EDCs were eliminated by the successful implementation of the method developed. 

The lack of aberrations in the processed images is similarly clear when extracting quantitative features from the subtracted data.  In Figure ~\ref{fig:quant}, we present one-to-one comparisons of normalized EDCs [Fig.~\ref{fig:quant}(a)] and MDCs [Fig.~\ref{fig:quant}(b)] from the He-I (red lines) and post-subtracted Ne-II (blue lines) spectra of Figures ~\ref{fig:misfit}(b) and (f), taken at identical cuts in momenta and energy.  We find that despite evidently greater measurement noise due to weaker signal intensity ($I_{Ne-II}/I_{He-I}$~$<$ 1$\%$), post-subtracted Ne-II EDCs and MDCs show nearly identical lineshapes to the He-I data, without any evident discontinuities or residual secondary peaks.  This can also be seen in the measured dispersion relations and MDC linewidths presented in Figures ~\ref{fig:quant}(c) and (d), respectively, which are extracted from Lorentzian fits to the He-I and post-subtracted Ne doublet data of Fig.~\ref{fig:misfit}.  We note that the Ne-I data exhibited an additional unrelated shoulder in the MDC lineshape and was excluded from the comparison of Fig.~\ref{fig:quant}(d) as a result.  The fitted results for the subtracted data again exhibit near-perfect agreement with the He-I behavior within measurement uncertainty.  Additionally, similar results were obtained when repeating the dispersion analysis on other cuts in momentum space, such as at the less intense band visible at positive $k_y$ in Fig.~\ref{fig:misfit}(b).  The excellent agreement across all measured and processed spectra demonstrates the practical efficacy of our algorithm at enabling quantitative analysis of ARPES spectra from doublet emission sources.

%[ uncovering the intrinsic ....]
%Similarly, fitted lorenMDCs taken both above and below the doublet energy separation show excellent qualitative agreement [Fig.~\ref{fig:quant}(b)].  

%Nevertheless, the quasiparticle peak is similar to that of the corresponding He-I EDC. Additionally, for bands with high velocity such as the ones displayed, MDC analysis is preferred. Our MDC analysis shows strong agreement in the quasiparticle peaks, underscoring the qualitative effectiveness of our algorithm. We then conducted an MDC analysis to quantify the performance. 

%After subtraction, the extracted dispersion relations exhibited near-perfect agreement [Fig.~\ref{fig:quant}(c),(d)] with the He-I dispersion relation. Similarly, the HWHM for post-subtracted Ne-II are in agreement with the He-I data. HWHM analysis was not conducted for the Ne-I data as we noted the presence of a shoulder [Fig.~\ref{fig:misfit}(b)] near $k=-0.7\si{\angstrom^{-1}}$ which would have affected the fitting. Such a shoulder is not present in the Ne-II data. Similar results were obtained when repeating the abovementioned dispersion and HWHM analyses on the less intense band, albeit with higher noise attributable to a weaker overall signal. This is expected as the two bands in [Fig.~\ref{fig:misfit}(b)] are symmetric. Our algorithm therefore shows excellent quantitative effectiveness in removing the $\beta$ signal. 

\section{Discussion}

It is worth noting that the simplified description of the doublet photoemission signal presented in Section II assumes that the binding-energy dependent EDC lineshapes of both the $\alpha$ and $\beta$ signals are identical.  In principle, however, the $\mathbf{k}$ position being sampled along an EDC on an ARPES image collected with a hemispherical analyzer varies with the kinetic energy in a fashion dependent on the exact measurement geometry~\cite{HufnerBook,doi:10.1063/1.5007226}.  A substantial variation in measured $\mathbf{k}$ between doublet line signals can in turn lead to differences in the intrinsic EDC lineshapes for the $\alpha$ and $\beta$ signals in the case of highly dispersive band features.  To evaluate the importance of this effect, we consider first the measured out-of-plane photoelectron momentum, given by

\begin{equation}
    k_{\perp} = \sqrt{2 m_e/\hbar^2}(V_0 + E_k \cos^2(\theta))^{1/2} ,
\end{equation}

where $\theta$ is the emission angle relative to normal and $V_0$ is the inner potential (typically $\sim$10 eV).  If we consider doublet lines of energy separation $\Delta E$ = 0.2 eV at normal emission, the absolute difference in $k_\perp$ is $\sim$ 0.01 $\si{\angstrom^{-1}}$, which for a typical perovskite lattice constant of $a=4 \si{\angstrom}$ is only $\sim0.5\%$ of the Brillouin zone. The in-plane momentum is 

\begin{equation}
    k_{\parallel} = \frac{1}{\hbar} \sqrt{2 m_e E_k} \sin(\theta) .
\end{equation}

In the case of $k_{\parallel}$, assuming a typical emission angle of $\theta = 25^{\circ}$, the absolute difference in $k_{\parallel}$ is $\SI{0.005}{\angstrom}^{-1}$, less than $0.3\%$ of a typical perovskite Brillouin zone.

The difference in sampled $\mathbf{k}$ for a typical doublet line is therefore negligible relative to the Brillouin zone, justifying our earlier treatment.  Similarly, we would anticipate that differences due to photoemission matrix element effects would also be minimal for the small photon energy differences observed here.

% and therefore also negligible.

%These calculations also support the results obtained and justify our assumption of having $k$ be approximately equal in equation (1).

\section{Conclusion}
We have created a simple method for effectively subtracting the undesired doublet signal's contribution to ARPES spectra in a way that is mathematically unique and insensitive to noise. This method is advantageous in that the process is more or less universal for any inert gas with higher energy doublet lines as it bypasses the need for specific VUV filters. Our quantitative analysis of this method's performance on the non-dispersive misfit compound $\textrm{(LaSe)\textsubscript{1.14}(NbSe\textsubscript{2})\textsubscript{2}}$ demonstrates its viability with regards to the determination of dispersion relations and quasiparticle lifetimes, particularly in 2D and weakly-dispersive non-2D systems. As having access to more photon energies is useful in general, labs with ready access to such multi-gas discharge lamps may benefit from this method at no additional expense. Our approach therefore unlocks numerous doublet emission lines for lab-based ARPES studies on quantum materials.

\begin{acknowledgments}
This work was conducted using the resources of Cornell University's Platform for Accalerated Realization, Analysis, and Discovery of Interface Materials (PARADIM), NSF Award Number 2039380. The authors declare no conflicting interests.
\end{acknowledgments}

\appendix
\counterwithin{figure}{section}

\section{Measurement of Source Emission Lines}
\begin{figure}
\includegraphics[width=\columnwidth]{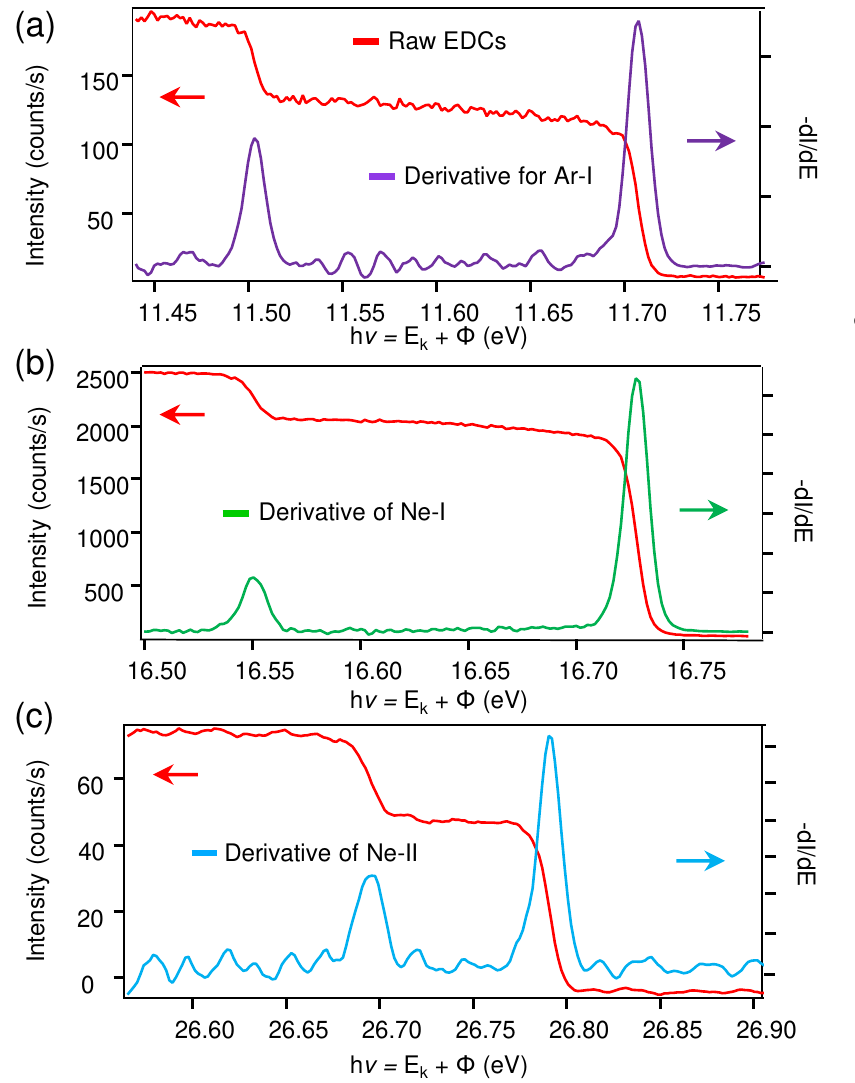}
\caption{\label{figure_linemethod}{Emission spectra and the negative of their derivatives measured with monochromator set to (a) Ar-I, (b) Ne-I, (c) Ne-II. The raw integrated EDCs for each line are shown in red, and the negative derivatives are plotted against the right axes in purple, green, and blue for Ar-I, Ne-I, and Ne-II respectively. These peaks were fit with Lorentzians to obtain the strong emission lines shown in Fig.~\ref{fig:problem}.}}
\end{figure}

The plasma discharge spectra shown in Fig.~\ref{fig:problem} were generated via angle-integrated photoemission measurements on a reference polycrystalline gold film, using the identical experimental setup and measurement conditions as for the ARPES data presented throughout the main text.  The gold reference sample was generated by e-beam deposition and cleaned via vaccum annealing before being transferred to the ARPES cryostat.  For photoemission measurements, the cleaned Au film was held at a temperature of ~6K under good UHV conditions ($<$ 7.5 x 10\textsuperscript{-11} Torr), with the electron analyzer set to a nominal energy resolution of 10 meV using He, Kr, Ar, and Ne gases to generate discharge emission.  In each case, raw integrated EDCs for each noble gas line were obtained, corresponding to the red curves on Fig.~\ref{figure_linemethod}. Ar-I, Ne-I, and Ne-II are colored in purple, green, and blue respectively. These curves were then offset by the work function of our analyzer, 4.32 eV, to give photon energies $h\nu = E_k + \Phi$. Taking the negative of the derivative for the integrated EDCs then produces the relative intensity and shape of the various strong emission lines after convolution with the detector energy resolution, as presented in Fig.~\ref{fig:problem}.

We also tabulate the lines shown in Fig,~\ref{fig:problem}(a) for our system in Table~\ref{table1} which should be a useful guide for future reference and for anyone who wishes to get an estimate of the various photon energies and relative intensities.
\begin{table}[t]
\begin{tabular}{cccc}
Line  & Normalized Intensity & Energy (eV) & R($I_\beta$/$I_\alpha$) \\ \hline
He-I  & 100\%                & 21.22      & -  \\ 
He-II & 0.51\%               & 40.83      & - \\ \hline
Kr-I  & 9.0\%                & 10.02      & - \\  \hline
Ar-I$\beta$  & 0.47\%               & 11.62     &  0.28  \\
Ar-I$\alpha$  & 1.7\%                & 11.83    &   \\  \hline
Ne-I$\beta$  & 6.8\%                & 16.67    &  0.22   \\
Ne-I$\alpha$  & 31\%                 & 16.85    &   \\  \hline
Ne-II$\beta$ & 0.30\%               & 26.82     & 0.40  \\
Ne-II$\alpha$ & 0.76\%               & 26.91    &  \\ 
\end{tabular}
\caption{\label{table1}Table of the emission lines shown in Fig.~\ref{fig:problem}, as measured via photoemission from a polycrystalline gold reference sample.  Normalized intensities are shown as a percentage of the He-I$\alpha$ count rate near $E_F$, and the energies are given as the measured $E_F$ in kinetic energy plus the analyzer work function.}
\end{table}

\section{Intensity Ratio Optimization}

\begin{figure}
\centering
\includegraphics[width=\columnwidth]{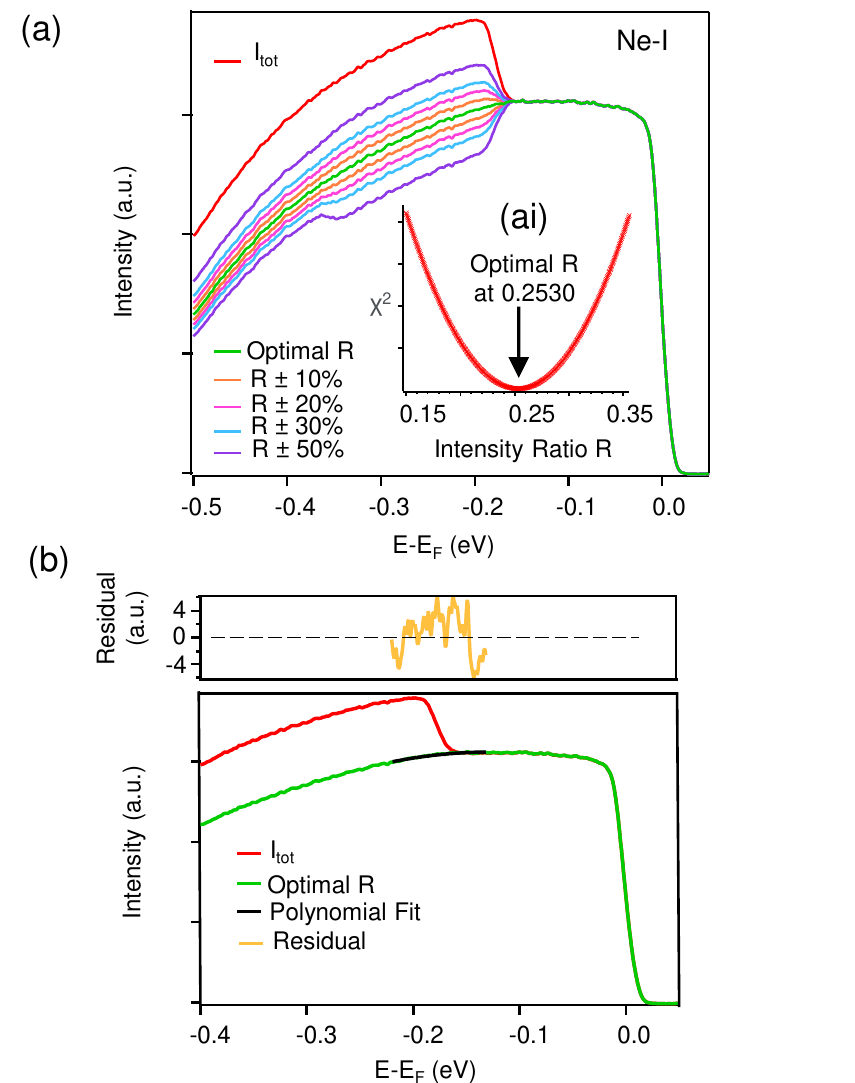} 
\caption{\label{fig:kink}{(a) Original and post-subtracted integrated EDCs extracted from the ARPES spectra of $\textrm{(LaSe)\textsubscript{1.14}(NbSe\textsubscript{2})\textsubscript{2}}$ misfit measured by Ne-I, which has a doublet separation of $\Delta E $ of 176 meV. Even slightly incorrect $R$ of $\pm$ 10 percent creates discontinuities at around $\Delta E$ below $E_F$. (ai) Chi-squared values when varying over R to find the optimal solution for which the discontinuity is minimized. (b) A closer examination of the red and green curves in (a), along with the smooth curve against which the residual from iterating $R$ is minimized. The yellow plot is the residual.}}
\end{figure}

Unlike $\Delta E$, $R$ can also vary between experiments using the same system on the same strong emission line due to differences in lamp pressure and monochromator settings. The biggest challenge is that there is no residual to minimize when varying $R$. Fortunately, even a slightly incorrect $R$ of $\pm 10\%$ of the optimal $R$ value yields a visible discontinuity in the integrated EDC at $\Delta E$ below the Fermi level. To illustrate this, we examined the integrated EDCs extracted from the ARPES spectra of $\textrm{(LaSe)\textsubscript{1.14}(NbSe\textsubscript{2})\textsubscript{2}}$ misfit measured by Ne-I. In Fig.~\ref{fig:kink}(a), we show the optimal solution from our algorithm in green, which corresponds to the solution using the optimal $R$ value. The orange, pink, blue, and purple curves are the solutions at non-optimal $R$, which create visible discontinuities in the EDC at $\Delta E$ = 176 meV below the Fermi level. Using this fact, we can fit a smooth cubic polynomial in a sub-window straddling the discontinuity, as shown in Fig.~\ref{fig:kink}(b), which then produces a residual against which the output of our subtraction algorithm can be minimized by varying $R$. This process of $R$ optimization is shown in Fig.~\ref{fig:kink}(ai), which plots the $\chi^2$ against the parameter $R$. Our optimal $R$ value is then the minimum of this optimization curve. We find empirically that choosing the width of this straddling sub-window centered around $\Delta E$ to be $\frac{\Delta E}{2}$ works well, and should be generally applicable for avoiding nearby bands which may complicate the process. An alternative method can be implemented utilizing smoothing followed by 2nd derivative analysis in this sub-window. We have chosen a cubic polynomial fit to avoid compounding noise when taking derivatives.

\section{Sensitivity to Noise}
\begin{figure*}
\includegraphics[width=\textwidth]{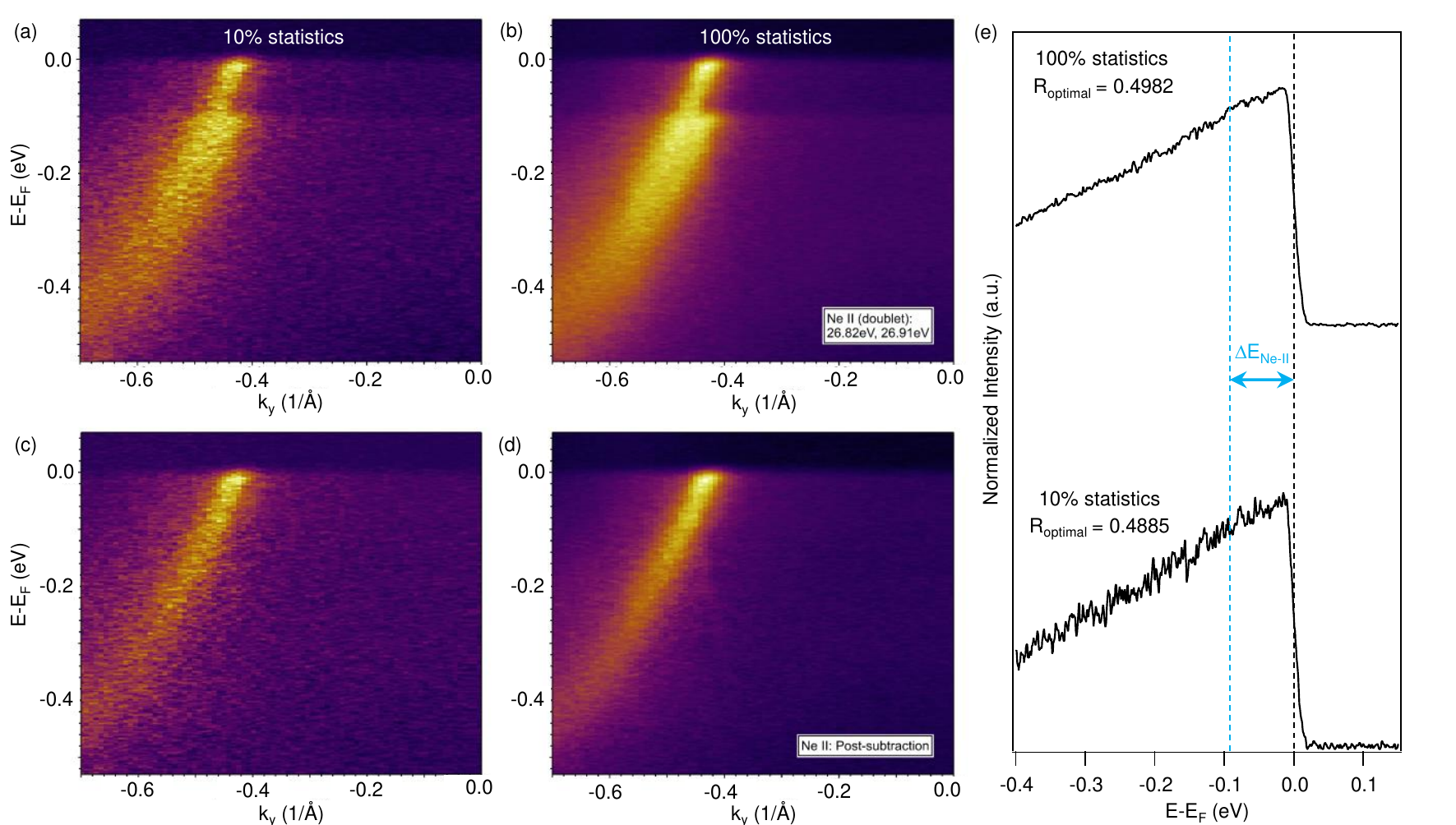}
\caption{\label{fig:noise}(a) Raw spectrum from 10$\%$ ARPES scan statistics taken with Ne-II. (b) Raw spectrum from 100$\%$ ARPES scan statistics taken with Ne-II. (c) Post-subtracted spectrum of (a). (d) Post-subtracted spectrum of (b). (e) EDCs for the 10$\%$ and 100$\%$ spectra integrated over a range of k=-0.6 $\si{\angstrom^{-1}}$ to k=0 $\si{\angstrom^{-1}}$}
\end{figure*}

Here we show that our algorithm performs well even on low ARPES statistics. Using the same data from the $\textrm{(LaSe)\textsubscript{1.14}(NbSe\textsubscript{2})\textsubscript{2}}$, but using 10\% and 100\% ARPES statistics, allows us to probe our algorithm's sensitivity to noise. In Fig.~\ref{fig:noise}(c) and (d), we show that the output spectra for 10\% and 100\% statistics are qualitatively identical. Further examination of the integrated EDCs [Fig.~\ref{fig:noise}(e)] also show the same quasiparticle peak and line shape. There is no kink in either the low or high statistics EDC at $\Delta E=90$ meV below the Fermi level corresponding to the doublet energy separation of Ne-II. Additionally, the two optimal $R$ values determined using the method in Appendix B exhibit a fractional difference of under 3\%, which attests to the algorithm's consistency even in the presence of worse signal-to-noise ratios. 

\section{Code}
An implementation of our algorithm written in Igor Pro 7 is available along with example Bi\textsubscript{2}Se\textsubscript{3} data shown in Fig.~\ref{fig:problem}(f) at a public Git repository \cite{GitHub}.

%\nocite{*}

\clearpage


\providecommand{\noopsort}[1]{}\providecommand{\singleletter}[1]{#1}%

\end{document}